\documentclass{article}

\usepackage{arxiv}

\usepackage[utf8]{inputenc} % allow utf-8 input
\usepackage[T1]{fontenc}    % use 8-bit T1 fonts
\usepackage{hyperref}       % hyperlinks
\usepackage{url}            % simple URL typesetting
\usepackage{booktabs}       % professional-quality tables
\usepackage{amsfonts}       % blackboard math symbols
\usepackage{nicefrac}       % compact symbols for 1/2, etc.
\usepackage{microtype}      % microtypography
\usepackage{lipsum}

\usepackage{float}
\usepackage{graphicx}
\usepackage{amssymb}
\usepackage{array}
\usepackage{longtable}
\usepackage{siunitx}
\usepackage{color}
\usepackage{multicol}
\usepackage{multirow}
\usepackage{comment}

\title{Learning Daily Calorie Intake Standard using a Mobile Game}

\author{
  Anik Das\\
  Department of Computer Science \& Engineering\\
  Chittagong University of Engineering \& Technology\\
  Chittagong, Bangladesh\\
  \texttt{anik.nicks@gmail.com} \\
 \And
 Sumaiya Amin \\
 Department of Computer Science \& Engineering\\
  Chittagong University of Engineering \& Technology\\
  Chittagong, Bangladesh\\
  \texttt{sumaiya.amin1997@gmail.com} \\
  \And
  Muhammad Ashad Kabir\\
  School of Computing and Mathematics\\
  Charles Sturt University\\
  Bathurst, NSW, Australia\\
  \texttt{akabir@csu.edu.au}\\
 \And
 Md. Sabir Hossain\\
 Department of Computer Science \& Engineering\\
  Chittagong University of Engineering \& Technology\\
  Chittagong, Bangladesh\\
  \texttt{sabir.cse@cuet.ac.bd} \\
  \And
  Mohammad Mainul Islam\\
  Verizon Media, Sunnyvale\\
  CA, USA\\
  \texttt{sujan.cse.cuet@gmail.com}\\
}

\begin{document}
\maketitle

% \begin{abstract}
% Mobile games can contribute to learning at greater success. In this paper, we have developed and evaluated a novel educational game, named FoodCalorie, to learn food calorie intake standard. Our game is aimed to learn calorie values of various traditional Bangladeshi foods and the calorie intake standard that varies with age and gender. We are the first in this field to perform an empirical study on women in Bangladesh to see how game-based learning can contribute to learn food calories. We further analyze and report the impact of participants' age, professions and smartphone proficiency levels on their learning experience and progression. Our study also conforms the finding of existing studies that game-based learning can enhance the learning experience.
% \end{abstract}

\begin{abstract}
Mobile games can contribute to learning at greater success. In this paper, we have developed and evaluated a novel educational game, named FoodCalorie, to learn food calorie intake standard. Our game is aimed to learn calorie values of various traditional Bangladeshi foods and the calorie intake standard that varies with age and gender. Our study conforms the finding of existing studies that game-based learning can enhance the learning experience.
\end{abstract}

% keywords can be removed

\section{Introduction}
\label{S:1}

Food is one of the basic needs of humans. For a healthy life, balanced food is necessary. Proper calorie intake can positively contribute to energy and well-being. Most people have no idea about the nutritional values of food, particularly the calories contained in each food. They either rely on ravishing food items without knowing what harm is caused, unconsciously leading themselves to fatal diseases, or they are simply deprived of proper nutrition~\cite{Shannon1994,Caamano2019,Hakim2016}.  

Improper nutrition can lead to health problems (e.g., obesity, malnutrition) caused by being overweight or underweight. A recent survey indicates that being obese or overweight may cause a devastating effect on health~\cite{WorldHealthOrganizationWHO2013}. Carrying excess fat leads to significant health hazards such as heart disease, stroke, and type 2 diabetes. It can also cause musculoskeletal issues such as osteoarthritis and some cancers, which might lead to undeniable impairment or even death. Malnutrition is a major cause of death in children and women. In addition to causing individual tragedies like maternal and child mortality, malnutrition results in excessive costs within the health care system through excess morbidity, increased premature delivery, and elevated risks of heart disease and diabetes~\cite{healthUNICEF}. School-age children who suffered from early childhood malnutrition have generally been found to have lower IQ levels, deficient cognitive functions, below average educational achievements, and greater behavioral problems~\cite{GranthamMcGregor1995}.
  
The global number of moderately or severely underweight girls and boys was 75 million and 117 million respectively in 2016. If post-2000 trends continue, the levels of child and adolescent obesity will surpass those for moderately and severely underweight youth from the same age group by 2022~\cite{WorldHealthOrganizationandothers2017}. 

A study, conducted by the Imperial College London and the World Health Organization (WHO), stated that Bangladesh is facing the ``dual burden" of both malnutrition and obesity~\cite{Islam2018}. According to a study, between 1975 and 2016, the weight problems (malnutrition and obesity) among boys in Bangladesh increased from 0.03\% to 3\%. Among girls, the rate increased from almost zero to 2.3\%~\cite{Khaled2017}.

The prevalence of obesity increased from 2.7\% to 8.9\% among women~\cite{Balarajan2009}. Childhood obesity is a particular public health concern for Bangladesh because children who are overweight or obese have a higher risk of becoming overweight or obese adults~\cite{Singh2008,Whitaker1997} and overweight adults are at increased risk for mortality and morbidity with obesity-associated chronic diseases, which are already a burden to the struggling health system in Bangladesh~\cite{Bhuiyan2013,Mirelman2012}.

At the same time, Bangladesh has the highest rate of malnutrition in the world. As per the Food and Agriculture Organization of the United Nations (FAO), among preschool-age children, a ratio of 54\% is stunted (which is greater than 9.5 million), whereas 56\% are underweight and more than 17\% are wasted further~\cite{FoodandAgricultureOrganizationoftheUnitedNationsFAO}. Almost half of Bangladeshi women suffer from chronic energy deficiency for a long run and research suggests that little improvement has been made in woman’s dietary conditions over the past twenty years. Proper knowledge of the food values is essential for living a healthy life. 

In Bangladesh, women hold the responsibility of cooking~\cite{Asaduzzaman2010,Barua2019,Charles1988,DeVault1991,Ekstrom1990,Furst1997}. They must be given proper knowledge of calories and nutrition in each food item. Only then they can ensure a nutritious and balanced meal for each family member considering their age, gender, and activity. However, the traditional teaching approach is not fruitful since it’s notoriously monotonous. A new technique needs to be introduced to overcome this challenging problem. 

\begin{comment}
A wide range of smartphone applications are available for increasing awareness of the nutrition value of raw food items (e.g., \cite{healthUNICEF}) such as raw meat, raw vegetables, eggs, and milk. A number of these apps have focused on changing the lifestyle of people suffering from diabetes, many operating by game-based learning approach (e.g., \cite{KamelBoulos2015}). Few of them have emphasized everyday exercise and daily nutrition intake for food in a specific country, for example, Korea (e.g., \cite{Lee2010}). One was developed only aiming at increasing fruit and vegetable intake among children~\cite{Baranowski2003}. None of these has focused on the Bangladeshi food (cooked) nutrition values with a goal to enhance knowledge of Bangladeshi women, especially about calorie requirements for different ages.
\end{comment}

In this paper, we have developed a game in which player will learn food calorie values. Specifically, we have designed our game, named “FoodCalorie”, for mobile phones and conducted an empirical study with 20 participants (Bangladeshi women) over two months period. Through pre- and post-intervention surveys, we studied how participants were acquired knowledge by playing the game. We have designed the game focusing on learning daily calorie intake standard for both male and female in different age, and calorie value of food items. Game has multiple levels, and in each level, a player is asked to choose food items for breakfast, lunch and dinner for a person with a specific age and gender. Food item and amount must be chosen according to age and gender. After the completion of each level, a reward is given. Thus, a player can learn calorie values of different food items, healthy meals for breakfast, lunch and dinner, and daily calorie intake standard for both male and female of different ages by playing our “FoodCalorie” game. 
%In this paper, we have proposed a game  in which women will learn food calorie values. Specifically, we designed a game for mobile phones called “FoodCalorie" and conducted a real-world evaluation of the game with 20 participants over a two-month period of time. Through pre- and post-intervention surveys, we studied how participants reacted to the game and how they were affected by playing. We have designed the game focusing on both age-wise calorie requirements and calorie value of food items. In the game, there are options for choosing food for three meals a day and designated by male or female. Food has to be chosen according to age. Game play has multiple levels and after the completion of each level, a reward is given. Playing “FoodCalorie" has helped our participants acquire knowledge about different food values, the progression difference among different occupations, and the acceptability of the game among women. Our aim is to have a realistic strategy, which can be spontaneously adopted by the women of Bangladesh. We have designed a mobile game in such a way to offer recreation with learning, since the women hardly get time to spend on recreation.

In particular, in this paper we have made the following three major contributions.
\begin{itemize}
    \item We have developed a mobile game to learn calorie intake standard. Our game is aimed to learn calorie values of various traditional foods of Bangladesh and the calorie intake standard that varies with age and gender.
    \item We have proposed and evaluated our novel approach: learning food value by game-based learning.
%    \item We have performed an empirical study related to knowledge learning, conceptual understanding, usefulness, and usability.
\end{itemize}

The remaining part of the paper is organized as follows. A comparative analysis of relevant research is discussed in ‘Related Work’ section. ‘Food Calorie Game’ section presents game architecture, interface and play rules. %The section ‘Empirical Evaluation’ presents the results of our empirical study; and 
finally, the last section concludes the paper and outlines the future work.

\section{Related Work}
\label{sec:relatedwork}

Game-based learning (GBL) refers to the use of miscellaneous types of games such as digital or non-digital games, simulations, and electronic games (e-games). It has one or more specific learning objectives along with teaching \& educational purposes~\cite{Wiggins2016, Tan2008, Tan2012, Cardinot2019, Wilson2013}.

Game-based learning has the ability to stimulate engagement, learning attitudes, and communication skills~\cite{Lui2018, Sprengel1994, Garris2002}. Additionally, it enriches the capability of understanding some critical concepts~\cite{Ganesh2014}. There are several studies which have proven success of game-based learning in primary, secondary and post- secondary education~\cite{Ke2008,Huizenga2009,Kolovou2010,Miller2011,Arnab2013,Annetta2009,Bourgonjon2010,Papastergiou2009,Lui2018}.

Games can develop healthy habits as demonstrated by one study~\cite{Brazendale2015}. Video games have shown useful for health and physical education ~\cite{Papastergiou2009}. Griffiths et al. surveyed several edutainment games for health and argued that video games can enrich the participants’ knowledge about food and influence them to lead a healthy life~\cite{Griffiths2013}. 

A game called ``Squire’s Quest!" was developed as part of a study~\cite{Baranowski2003}. The study showed that games can enhance the fruit, juice, and vegetable intake of children.

The games discussed above were aimed at teaching different groups of people about healthy foods and nutrition values for different kinds of food. None of the above-mentioned games has considered food items with specific unit of measurement. Most importantly, none of the above aims to educate daily calorie intake standard of a person with different ages and gender. Compared to the above research, in our game, we consider food items that are appropriate to the specific meal such as breakfast, lunch, and dinner. We have also considered daily calorie intake standard for individual person considering age and gender. We are the first to study this game-based learning in Bangladesh. %A comparative summary between our study and other related studies is presented in Table~\ref{tab:literatureReview}.

\section{FoodCalorie Game}
\label{sec:proposedgame}
In this section, we will discuss our game architecture and interface in detail, including ratings and rules for achieving each level of the game.

\subsection{Game Architecture}
\begin{figure}[!htbp]
\centering\includegraphics[width=.75\textwidth]{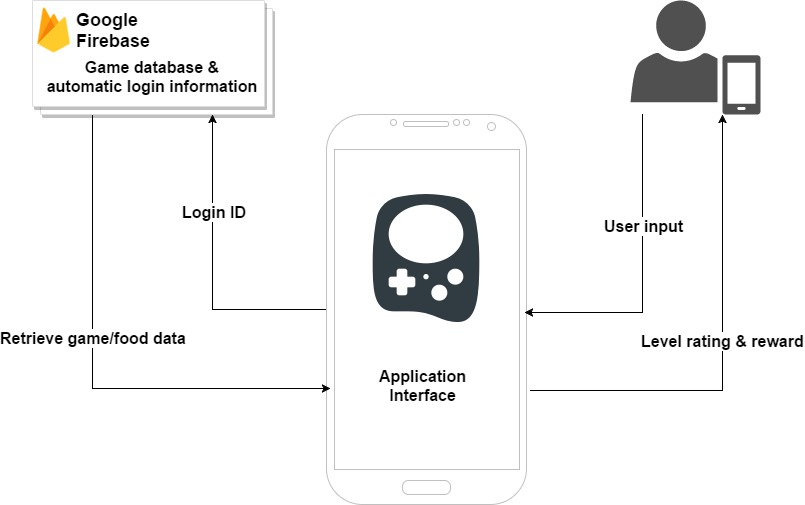}
\caption{Architecture of our FoodCalorie game}
\label{fig:architecture}
\end{figure}

Figure~\ref{fig:architecture} shows the game architecture. Our game uses Google Firebase\footnote{https://firebase.google.com/} where food and calorie information is stored. Thus, an internet connection is required for playing our game. When the game is installed for the first time, a Login ID is generated by the Firebase, which is stored in the user's mobile phone for automatic login on future game play. When a user opens the game, the stored Login ID is used to authenticate the user in the Firebase and retrieve game data to display on the game application interface.

\subsection{Game Interface}
\begin{figure}[!htbp]
\centering\includegraphics[width=.75\textwidth]{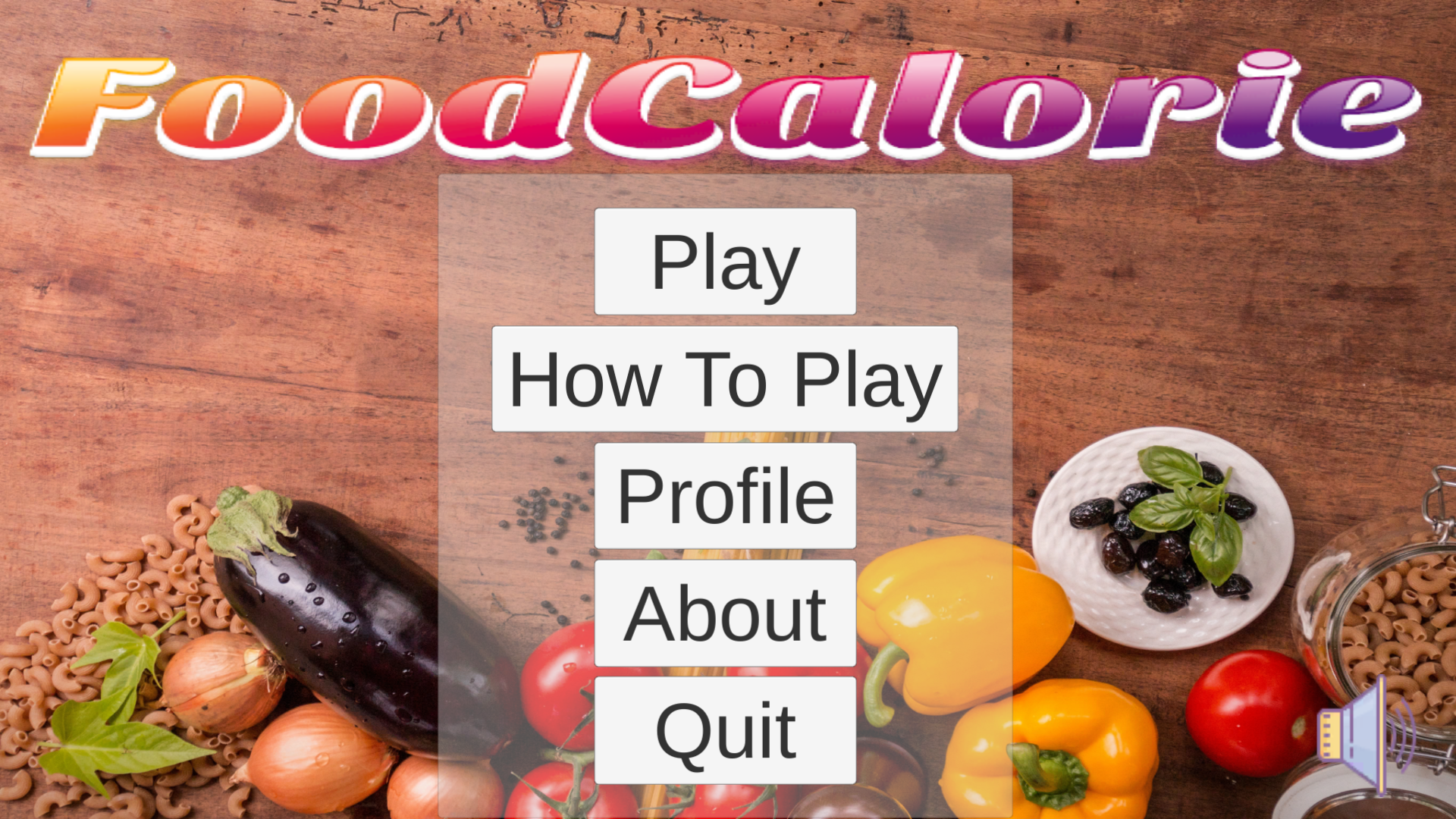}
\caption{Home screen of our game}
\label{fig:homescreen}
\end{figure}

\begin{figure}[!htbp]
\centering\includegraphics[width=.75\textwidth]{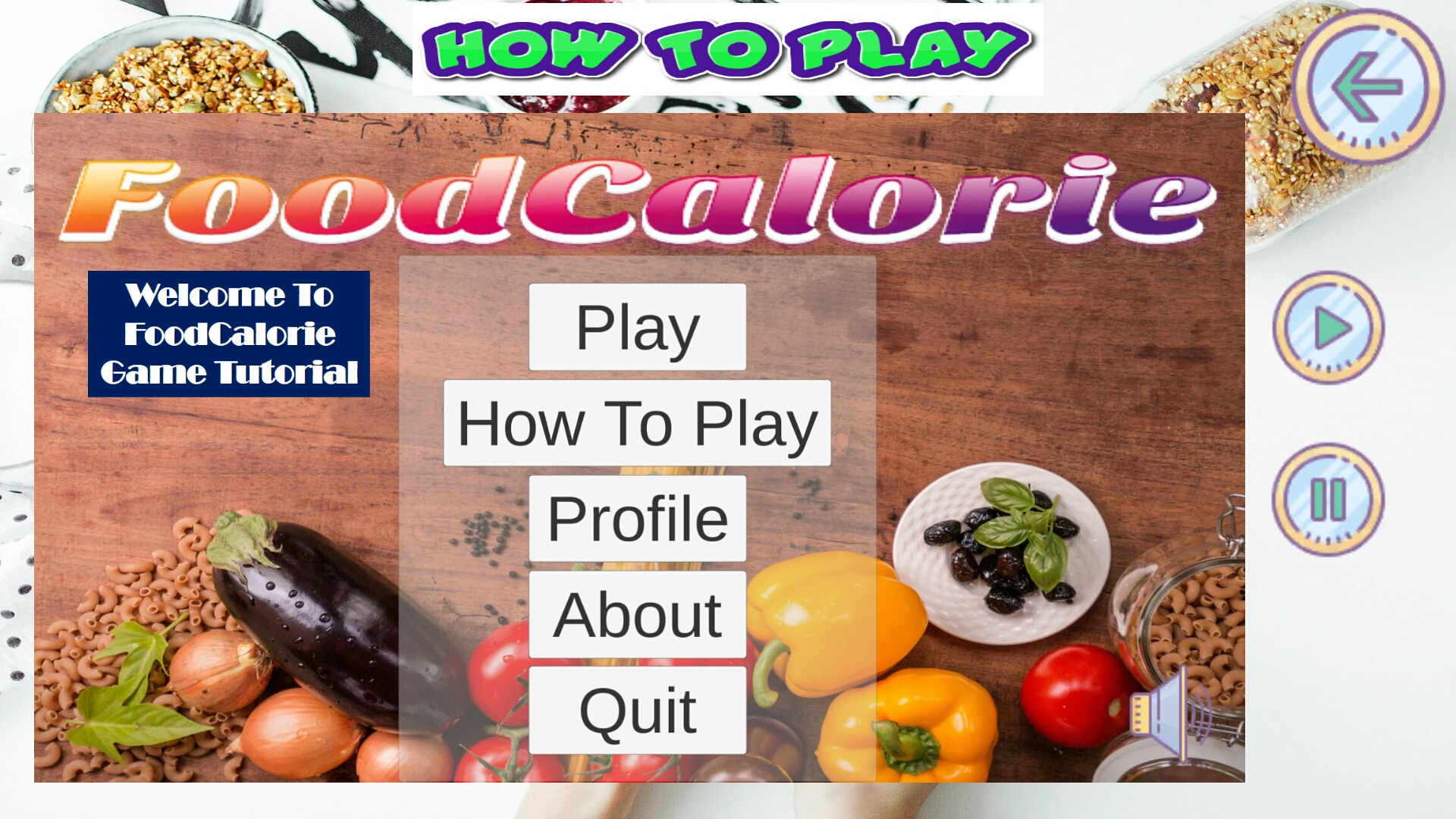}
\caption{How to play screen}
\label{fig:demoscreen}
\end{figure}

At the very beginning, the home screen will be shown (Figure~\ref{fig:demoscreen}). This screen shows the user the following options: Play – start to play the game; How To Play – short demo video on playing this game; Profile – total number of levels with how many levels tried and passed; About – information related to the creation of the game; and Quit – to exit the game (see screenshot Figure~\ref{fig:homescreen}).

%\subsubsection{Main Game Structure}

A person’s daily calorie intake requirements vary by gender and age. It also has a strong relationship with physical activity. For instance, a sedentary person usually requires fewer calories than a moderately active person does~\cite{UnitedStatesDepartmentofAgricultureUSDA2002}. In our game, we have used the average calorie value of sedentary and moderately active levels (assuming that the person does some physical activities). For calorie calculation, we have used data from a reputable health website~\cite{Medindia}. 

\begin{figure}[!htbp]
\centering\includegraphics[width=.7\textwidth]{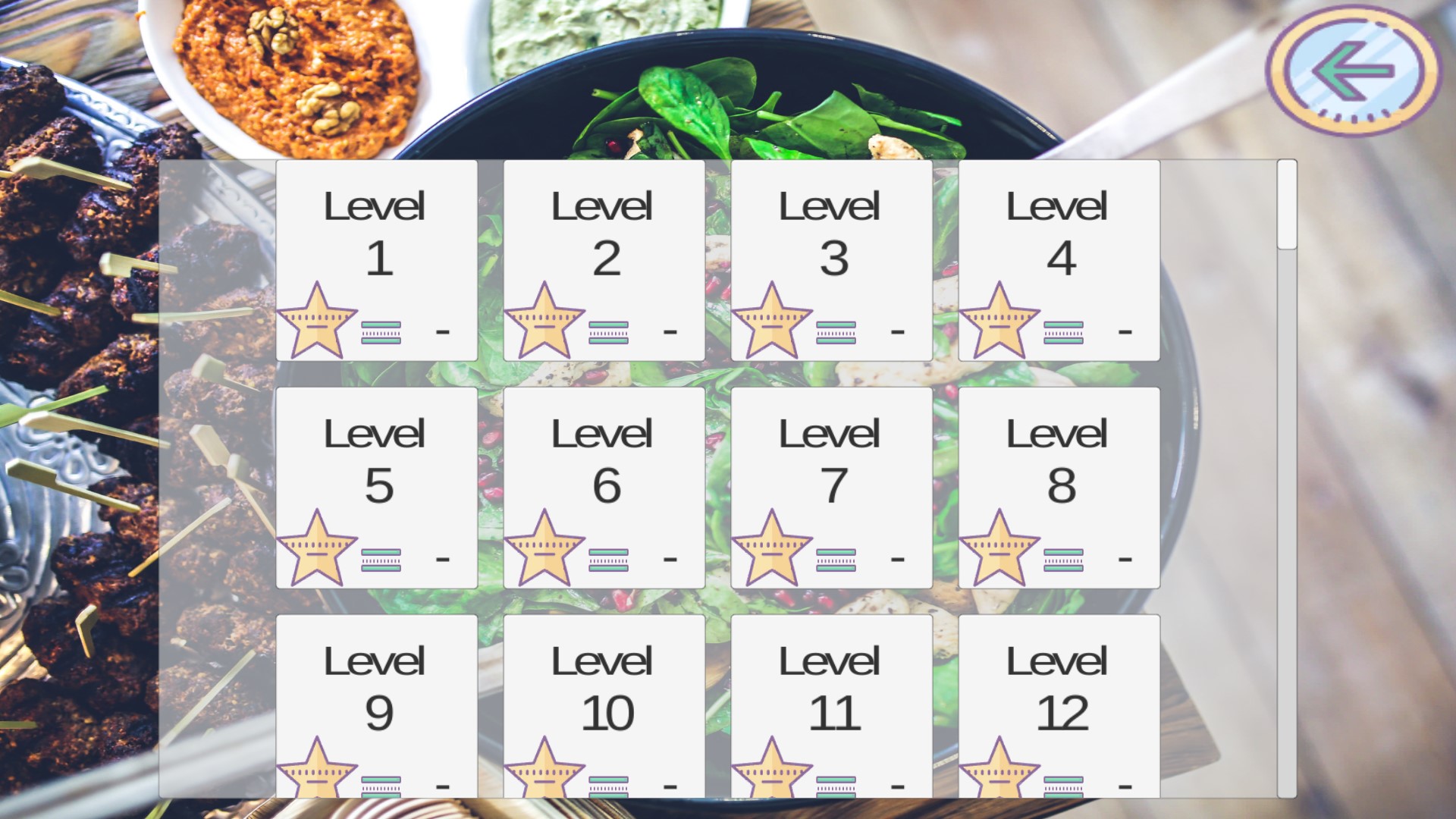}
\caption{Level division of our game}
\label{fig:levelscreen}
\end{figure}

In this game, users are required to select food that provides appropriate calorie intake for a person with a particular gender and age. Our game has 96 levels (each level is for a particular age) starting from 3 years to 99 years. We did not consider 1- and 2-year-old children as they have special meal requirements. Figure~\ref{fig:levelscreen} shows the level screen of our game. At each level, foods need to be chosen for a person at a particular age. The meals are planned as – breakfast, lunch, and dinner. The meals are planned as – breakfast, lunch, and dinner. There are six individual identical windows on each level. The first three windows are for male breakfast, lunch, and dinner and the next three are for female of a particular age.
%There are six individual identical windows on each level. The first three windows are for (male) breakfast, lunch, and dinner and the next three are for (female) breakfast, lunch, and dinner of a particular age.

We have included seventy-two food items in this game that are classified into seven types. Eight items are rice (e.g., plain rice, biriyani, khichuri, etc.); five items are bread (e.g., roti, slice bread, naan, etc.); thirty-two items are curry (e.g., meat curry, fish curry, vegetable fry, etc.); eleven items are fruit (e.g., pineapple, jack fruit, mango, etc.); six items are dessert (e.g., kheer, pudding, halwa, etc.); four items are dairy (e.g., eggs and milk); and six items are classified as other (including fast-food, e.g., burger, haleem, etc.). In our game window, the user needs to select three food items out of six given items chosen randomly from 72 food items.

Every country has its own tradition of food. In Bangladesh, rice and bread are considered the main food staples for breakfast, lunch, and dinner. Thus, we have designed our food selection pool in such a way that there are always two items from rice and bread types. The remaining four items will be chosen randomly from the other five assortments.

\begin{comment}
Figure~\ref{fig:beforeselectoin} shows the food selection window of our game. In this window, a user needs to choose 1 to 3 items from the six items shown, using a drag-and-drop function. It also has an option to specify the quantity of the added item(s) by increasing (pressing the '+' button) or decreasing (pressing the '-' button) the units. Figure~\ref{fig:afterselection} shows the game after the selection of food items. As appropriate, in our game, we have considered four types of units to measure food: 100 gm (for rice, dessert, etc.), piece (for bread, fruits, etc.), glass (for milk), and cup (for some curries). The total calorie count for each item for the specified quantity will be calculated and shown in the window. At the top of the window, we display the required daily calorie total (i.e., breakfast, lunch, and dinner together) for a person with a specific gender and age. This needs to be considered while selecting items for breakfast, lunch, and dinner. The goal is to ensure that the total calorie count of the selected food does not exceed the daily limit.

\begin{figure}[!htbp]
\centering
\includegraphics[width=.75\textwidth]{fig-20.jpg}
\caption{Before selection of food items (for female age 33 breakfast)}
\label{fig:beforeselectoin}
\end{figure}

\begin{figure}[!htbp]
\centering
\includegraphics[width=.75\textwidth]{fig-21.jpg}
\caption{After selection of food items (for female age 33 breakfast)}
\label{fig:afterselection}
\end{figure}

\end{comment}

\subsection{Rating and Level Passing Rules}
\begin{figure}[!htbp]
\centering\includegraphics[width=.75\textwidth]{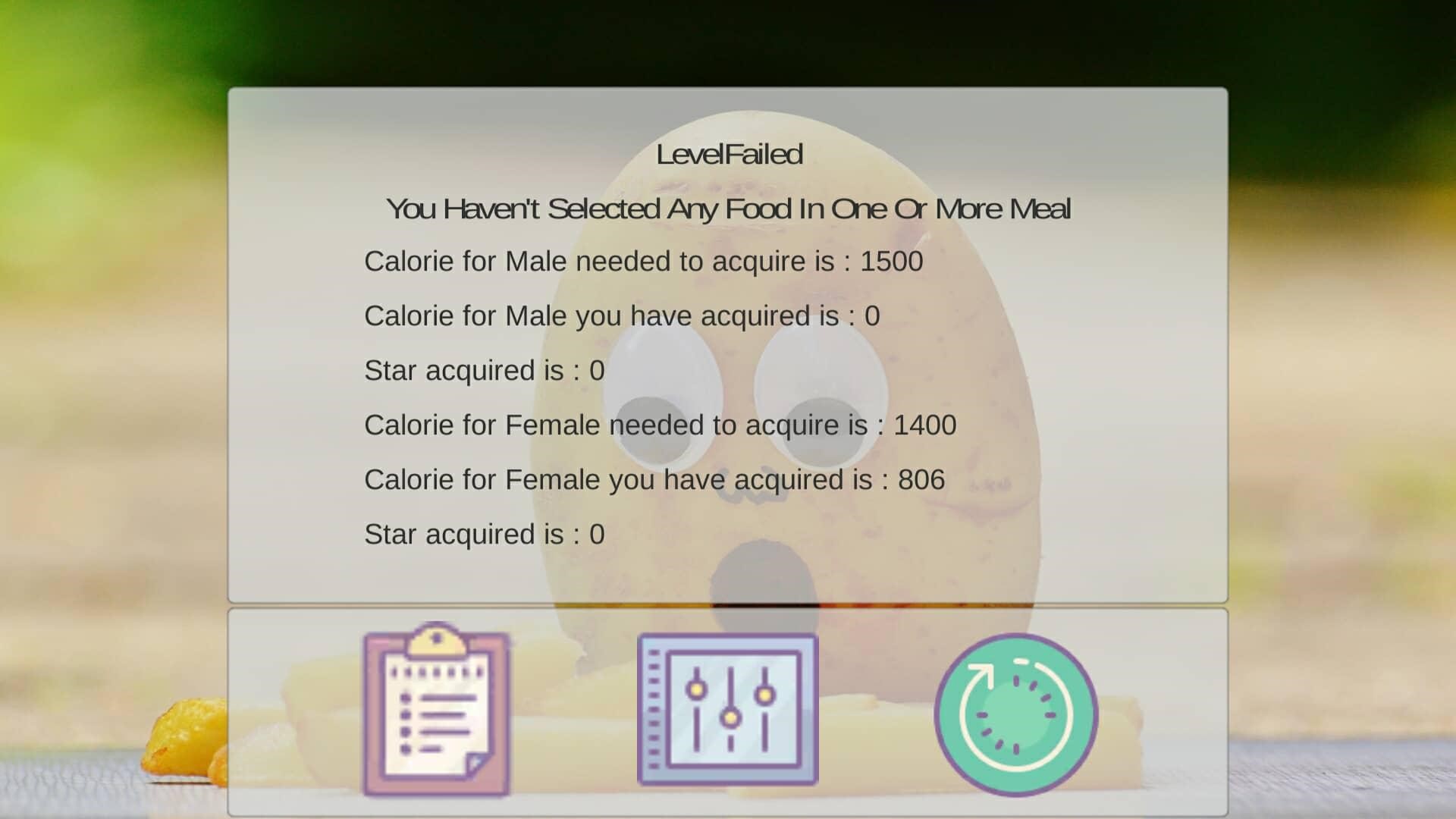}
\caption{Window for showing the result}
\label{fig:levelsummary}
\end{figure}

A level comprises six selection windows where a user selects breakfast, lunch, and dinner for a male and female. At the end of each level, a summary result is shown (Figure~\ref{fig:levelsummary}). 

Based on the \textit{acquired calories} (corresponding to the selected meals) for male or female we award zero, one, two, or three stars to the game player as shown below:

\begin{itemize}
\item If the \textit{selected calorie} $\leq$ \textit{required calorie}+5 and \textit{selected calorie} $\geq$ \textit{required calorie}$-5$, then the player will get 3 stars.

\item If the \textit{selected calorie} $\leq$ \textit{required calorie}+10 and \textit{selected calorie} $\geq$ \textit{required calorie}$-10$, then the player will get 2 stars.

\item If the \textit{selected calorie} $\leq$ \textit{reqquired calorie}+20 and \textit{selected calorie} $\geq$ \textit{reuired calorie}$-20$, then the player will get 1 star.

\item otherwise no star will be awarded
\end{itemize}

\section{Conclusion and future work}
\label{sec:conclusion}

In this paper, we have developed and evaluated a game for learning calorie intake standard. We have conducted an empirical study to investigate how this game can enhance food calorie knowledge of the players, in particular, women in Bangladesh. The results of our empirical study show that the calorie intake knowledge of all participants improved above 85\% by playing our game. The t-test result of our study shows that the knowledge enhancement of participants (with p value $<<$ 0.001) was significant. The game also enhanced the conceptual understanding of all participants above 91\%. 94\% of participants found this game is useful for learning healthy food habits. 95\% of participants have rated our game 4 stars and above. We have also found that participants who are a doctor in the profession have achieved higher learning experience compared to other professions and the participants with the highest smartphone proficiency level have a higher progression rate.

As future work, our aim is to run a longitudinal study to investigate how this game can gradually encourages players/women to make better food choices for their family. Besides, we will explore how this game change the eating habits of individual players over the time.

%In this paper, we have proposed a food calorie game and studied the impact of this game on teaching women in Bangladesh about food, nutrition, and healthy meal preparation. Our game is aimed to teach proper calorie intake requirements and the caloric values of various traditional foods in South-Asian countries, in particular, Bangladesh. We are the first in this field to perform an empirical study on women in Bangladesh to see how gamification can contribute to acquiring knowledge about food calories (based on age and gender particularly). We have exhibited knowledge enhancement of our participants considering several factors such as occupation, age, and smartphone proficiency level. Further, we have demonstrated the progression of our participants for establishing the fact that our proposed game can contribute to learning. For evaluating the acceptance of our game, we have studied user satisfaction, which identifies some suggestions from participants to improve this game further.
%As future work, we will develop the food items for the ease of users. We will also work on the graphics and display for making the game more attractive, along with working on the fields where users are currently facing difficulty. In the long term, we want to study whether the game leads to changes in women’s knowledge of food and nutrition and if it encourages them to make better food choices for their family from a healthy mindset.

% \section*{Data Availability Statement}
% The data that support the findings of this study are available from the corresponding author upon reasonable request.

\section*{Acknowledgement}
Author Anik Das and author Sumaiya Amin have contributed equally to this paper. All the authors have agreed to acknowledge this information.

\bibliographystyle{unsrt}  
\bibliography{main}

\end{document}